\documentclass[a4paper]{jpconf}
\usepackage{graphicx}

\begin{document}
\title{Heavy Hadrons in Dense Matter}

\author{Laura Tolos$^{1,2}$, Carmen Garcia-Recio$^3$, Carlos Hidalgo-Duque$^4$, Juan Nieves$^4$, Olena Romanets$^5$,  Lorenzo Luis Salcedo$^3$ and Juan M. Torres-Rincon$^6$}

\address{$^1$Instituto de Ciencias del Espacio (IEEC/CSIC), Campus UAB, 
Carrer de Can Magrans s/n, 08193 Cerdanyola del Valles, Spain}

\address{$^2$Frankfurt Institute for Advanced Studies, Johann Wolfgang Goethe University, Ruth-Moufang-Str. 1,
60438 Frankfurt am Main, Germany} 

\address{$^3$Departamento de F{\'\i}sica At\'omica, Molecular y Nuclear, and Instituto Carlos I de F{\'i}sica Te\'orica y Computacional,
Universidad de Granada, E-18071 Granada, Spain}

\address{$^4$Instituto de F{\'\i}sica Corpuscular (centro mixto CSIC-UV),
Institutos de Investigaci\'on de Paterna, Aptdo. 22085, 46071, Valencia, Spain}

\address{$^5$KVI, University of Groningen, Zernikelaan 25, 9747AA Groningen, The Netherlands}

\address{$^6$Subatech, UMR 6457, IN2P3/CNRS, Universit\'e de Nantes, \'Ecole de Mines de Nantes, 4 rue Alfred Kastler 44307,
Nantes, France}

\ead{tolos@ice.csic.es}

\begin{abstract}
We study the behavior of dynamically-generated baryon resonances with heavy-quark content within a unitarized coupled-channel theory in matter that fulfills heavy-quark spin symmetry constraints. We analyze the implications for the formation of charmed mesic nuclei and the propagation of heavy mesons in heavy-ion collisions from RHIC to FAIR.
\end{abstract}

\section{Introduction}
\label{intro}

The study of the properties of hadrons under extreme conditions is one of the main research activities of several experimental programs and, in particular, of the forthcoming FAIR  (Germany) project. The aim is to move from the light-quark sector to the heavy-quark domain and to face new challenges where charm and new symmetries, such as heavy-quark symmetries, will play a significant role.

One of the primary goals is to understand the nature of newly discovered states and, in particular, baryonic states with charm and bottom degrees of freedom. In that respect, approaches based on unitarized coupled-channel dynamics have shown a tremendous success in the past. Recently, a unitarized coupled-channel scheme that incorporates  heavy-quark spin symmetry (HQSS) \cite{Isgur:1989vq} constraints has been developed  \cite{GarciaRecio:2008dp,Gamermann:2010zz,Romanets:2012hm,GarciaRecio:2012db,Garcia-Recio:2013gaa,Tolos:2013gta,Garcia-Recio:2015jsa}. HQSS is a proper symmetry of the strong interaction that appears when the quark masses become larger than the typical confinement scale.  Furthermore, nuclear medium corrections have been implemented \cite{Tolos:2009nn,GarciaRecio:2010vt,GarciaRecio:2011xt} to study the properties of the newly discovered heavy baryonic states in dense matter and their influence on heavy mesons in nuclear matter and nuclei.

In this work we study the properties of heavy hadrons in dense matter.  We aim at investigating nuclear medium effects on dynamically-generated heavy baryonic resonances and the consequences for the formation of charmed mesic nuclei as well as the propagation of heavy mesons under extreme conditions.


\begin{table}
\caption{Compositeness of  $\Lambda$ states in the charm (left table) and bottom (right table) sectors (taken from Ref.~\cite{Garcia-Recio:2015jsa}). The masses ($M_R$) and widths ($\Gamma_R$) are given in ${\rm MeV}$.}
\label{fig:reso}
\smallskip\noindent
\resizebox{\linewidth}{!}{%
{\begin{tabular}{cccccccrr}
\hline
State & $J^P$  & $M_R$ & $\Gamma_R$ & 1$-Z$ & Channel & $|g_i|$  & $X_i$
\\ 
\hline
$\mathbf{\Lambda_c(2595)}$  & $\frac{1}{2}^- $ &  $2619$ &   $ 1.2 $  & $0.878$ & 
$\pi   \Sigma_c   $ &  $0.31$  & $-0.012$  \\
&&&&& 
${D N} $ &  $3.49$  & $ {0.275}$  \\
&&&&&
 ${ D^* N} $ &  $5.64$  & $ {0.465}$   \\
\\
 \hline\hline
$\mathbf{\Lambda_c(2595)}$  & $\frac{1}{2}^- $ & $2617$ &   $90$ &  $ 0.401$  &
 ${\pi \Sigma_c} $ & $ 2.36$  & $ {0.325}$ \\
&&&&&
 $D           N      $ & $ 1.64$  & $ 0.027$  \\
&&&&&
 ${D^* N}$ &  $1.43$  & $ 0.024$  \\ \\
\hline\hline

$\mathbf{\Lambda_c(2625)}$  & $\frac{3}{2}^- $ &  $2667$ &   $55$ & $0.365$ &
 ${\pi \Sigma_c^*} $ &  $2.19$  & $ {0.268}$  \\
&&&&&
${ D^* N }$ &  $2.03$ &  $ {0.057}$  \\ \\\hline
\end{tabular}
}~~{\begin{tabular}
{cccccccrr}
\hline
State & $J^P$ &  $M_R$ & $\Gamma_R$ & 1$-Z$ & Channel & $g_i$~ & $X_i$~~\\\hline
$\mathbf{\Lambda_b(5912)}$  & $\frac{1}{2}^- $ & $5878$ &   $ 0$ & $ 0.956$ &
 $\pi   \Sigma_b   $ & $ 0.04  $ & $ 0.000$  \\
&&&&&
 ${ \bar{B} N}$  & $-4.55  $ & $ {0.205}$  \\
&&&&&
 $ { \bar{B^*} N} $  & $-7.70  $ & $ {0.539}$   \\
\\
\hline\hline

$\mathbf{\Lambda_b(5912)}$ & $\frac{1}{2}^- $ & $5949$ &   $ 0$ &  $0.865$ &
 $ {\pi \Sigma_b }$  & $ 1.31 $ & $ {0.698}$  \\
&&&&&
 $ { \bar{B} N} $  & $-2.90 $ & $ {0.096}$ \\
&&&&&
 $ {\bar{B^*} N} $  & $ 1.91 $ & $ 0.038$   \\
\\
\hline\hline

 $\mathbf{\Lambda_b(5920)}$ & $\frac{3}{2}^- $ & $5963$ &   $ 0 $ &  $ 0.818$ &
 $ { \pi \Sigma_b^*} $  & $ 1.54  $ & $ {0.581}$  \\
&&&&&
 $ { \bar{B^*} N }$  & $ 4.16  $ & $ {0.185}$  \\ \\
 \hline
\end{tabular} 
}}
\end{table}

\section{Compositeness of dynamically generated heavy  $\Lambda$ states}
\label{spec}

Heavy baryonic states are dynamically generated by the scattering of mesons and baryons within a unitarized coupled-channel approach. In this work we employ a model that explicitly incorporates HQSS~\cite{Isgur:1989vq}. HQSS predicts that all types of spin interactions involving heavy quarks vanish for infinitely massive quarks, thus, connecting vector and pseudoscalar mesons containing heavy quarks. Furthermore,  chiral symmetry fixes  the lowest order interaction between Goldstone bosons and other hadrons by means of the Weinberg-Tomozawa (WT) term. This predictive model  includes all basic hadrons (pseudoscalar and vector mesons, and $1/2^+$ and $3/2^+$ baryons) and it reduces to the WT interaction in the sector where Goldstone bosons are involved while incorporating HQSS in the sector where heavy quarks participate. This scheme is justified in view of the reasonable outcome of the SU(6) extension in the three-flavor sector \cite{Gamermann:2011mq} and on a formal plausibleness in the vector-meson exchange picture of the interaction in the heavy pseudoscalar meson-baryon sector.

The extended WT model with HQSS constraints is used as the kernel of  the on-shell Bethe-Salpeter equation in coupled channels so as to calculate the scattering amplitudes. The poles of the scattering amplitudes are the dynamically-generated heavy baryonic resonances. In this work we present results in the sector with heavy (charm/bottom) (H), strange (S) and isospin (I) content such as $H=1,S=0, I=0$, where the $\Lambda$ states are found \cite{Garcia-Recio:2015jsa}.

We study the generalized Weinberg's sum rule \cite{Weinberg:1962hj} to estimate the importance of the different meson-baryon channels for the generation of the $\Lambda$ states. In Table~\ref{fig:reso} we show the spin-parity ($J^P$), mass ($M_R$) and width ($\Gamma_R$) of the different charmed and bottomed $\Lambda$ states, together with the absolute value of the coupling to the dominant meson-baryon channels ($|g_i|$).  The quantity $\sum_i X_i = 1-Z$ represents the {\em compositeness} of the
hadronic state in terms of all the considered channels, and $Z$ is referred to
as its {\em elementariness}. A small value of $Z$ indicates that the state is well
described by the contributions explicitly considered, namely, $s$-wave
meson-baryon channels, while a larger value of $Z$ indicates that, for
that state, significant pieces of information are missing in the model.

We obtain two $J^P=1/2^-$ and one $J^P=3/2^-$ $\Lambda$ states. We find that the $\Lambda$ states
 which are bound states (the three $\Lambda_b$) or narrow resonances (one $\Lambda_c(2595)$) are well described as molecular
 states composed of $s$-wave meson-baryon pairs. The $1/2^-$ wide $\Lambda_c(2595)$ as well as the $3/2^-$
 $\Lambda_c(2625)$ states display smaller compositeness. With respect to the detailed composition of the states, we find that the first
$\Lambda(1/2^-)$ states of each flavor couple strongly to
pseudoscalar-$N$ and vector-$N$ channels. For the second $\Lambda(1/2^-)$ states, the main observation is its
sizable coupling to the lightest channels  $\pi \Sigma_c(\Sigma_b)$. Another observation is the similar structure of the second
$\Lambda(1/2^-)$ and $\Lambda(3/2^-)$ states, which appear as
HQSS or spin-flavor partners.

\section{Charmed mesons in nuclei}
\label{medium}

\begin{figure}
\begin{center}
\includegraphics[height=0.4\textwidth,width=5cm,angle=-90]{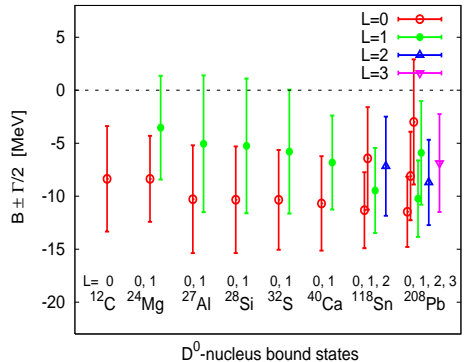}
\caption{$D^0$-nucleus bound states (taken from Ref.~\cite{GarciaRecio:2010vt}). }
\label{fig:medio}
\end{center}
\end{figure}

The in-medium modifications of the dynamically-generated $\Lambda$ states (and others) have important consequences on the properties of heavy mesons in matter. In particular, the properties of open-charm mesons in matter have been object of recent analysis due to the consequences for charmonium suppression. Moreover, the existence of charmed meson bound states in nuclei have been predicted  in $^{208}$Pb \cite{Tsushima:1998ru}, using an attractive $D$ and $\bar D$ -meson potential in the nuclear medium within a quark-meson coupling (QMC) model. The experimental observation of these bound states, though, can be problematic since their widths could be very large compared to the separation of the levels. 

Within our model, we obtain that the $D^0$-nucleus states are weakly bound (see Fig.~\ref{fig:medio}), in contrast to previous results using the QMC model. Moreover,  those states have significant widths \cite{GarciaRecio:2010vt}, in particular, for $^{208}$Pb. Only $D^0$-nucleus bound states are possible since the Coulomb interaction forbids the formation of bound states for $D^+$ mesons. As for $\bar D$ mesons  in nuclei, not only $D^-$ but also $\bar{D}^0$ bind in nuclei  \cite{GarciaRecio:2011xt} . 

The information on bound states is very valuable for gaining insight into the meson-nucleus interaction at the PANDA experiment at FAIR. Nevertheless, the experimental observation of $D$ and $\bar D$-meson bound states is a difficult task. Open-charm mesons with high momenta would be produced in antiproton-nucleus collisions at PANDA and it is a challenge to bind them in nuclei \cite{GarciaRecio:2010vt}.

\section{Heavy-meson propagation in hot dense matter}

Information on the properties of heavy mesons in matter can be also achieved by analyzing the heavy-meson propagation in the hot and dense medium created in heavy-ion collisions \cite{Ozvenchuk:2014rpa,Song:2015sfa}. The heavy-meson propagation can be studied by means of solving the corresponding Fokker-Planck equation. The two relevant quantities to be determined are the drag ($F_i$) and diffusion coefficients ($\Gamma_{ij}$) of heavy mesons in hot dense matter.  These are obtained from an effective field theory that incorporates both the chiral and HQSS in the meson \cite{Abreu:2011ic} and baryon sectors \cite{Tolos:2013kva,Torres-Rincon:2014ffa}.

\begin{figure}
\begin{center}
\includegraphics[width=0.4\textwidth,height=5cm]{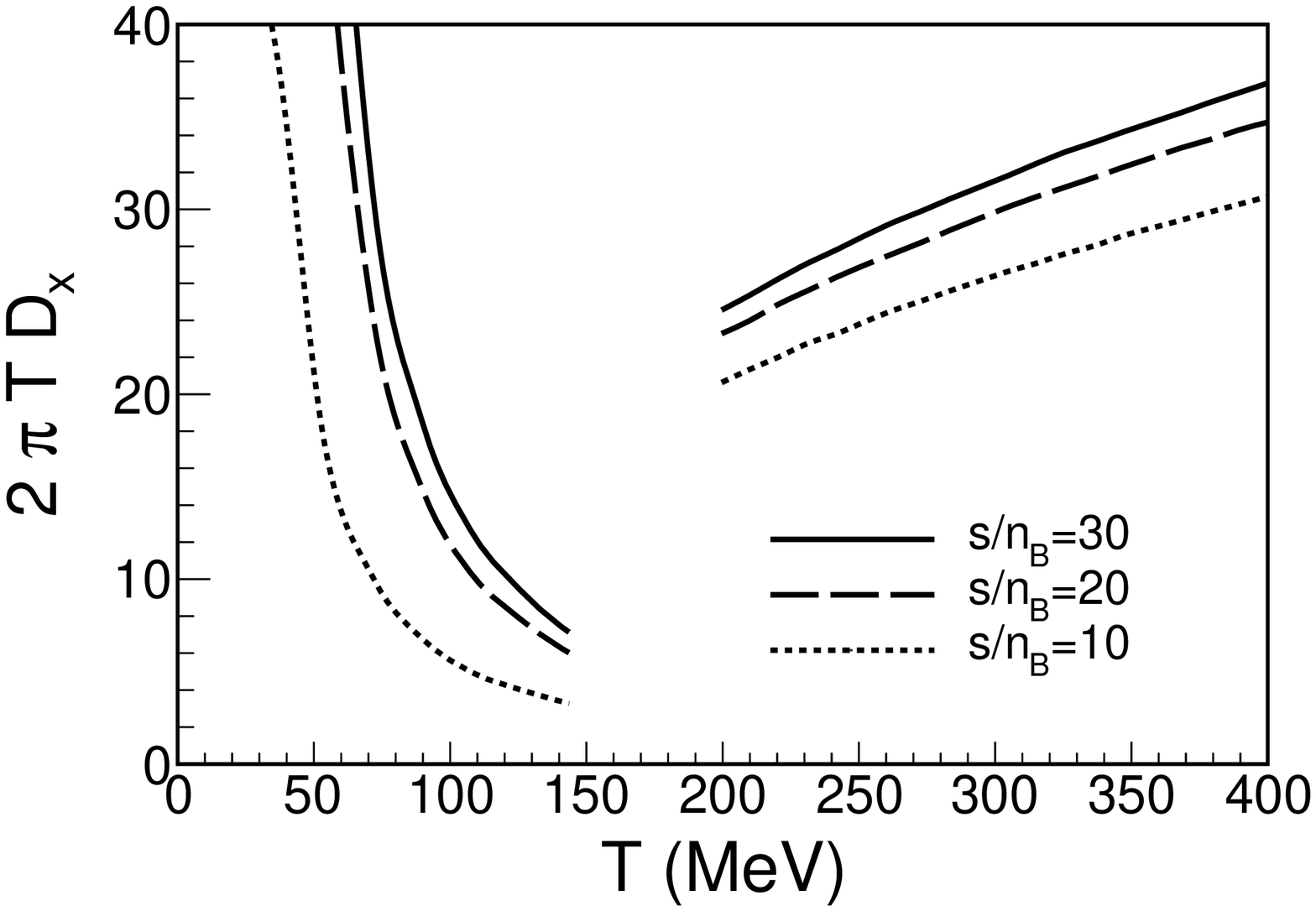}
\includegraphics[width=0.38\textwidth, height=5cm]{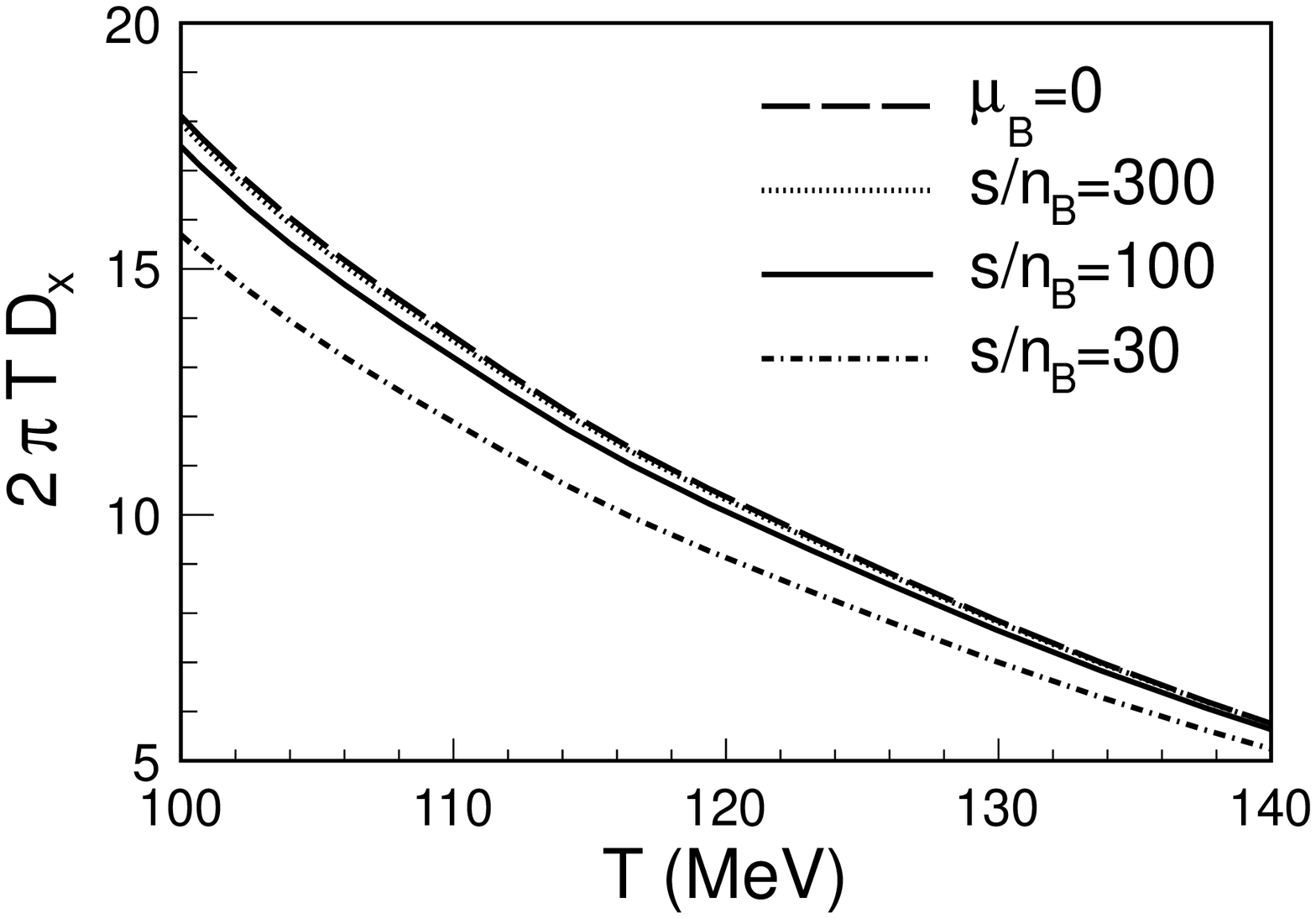}
\caption{The coefficient $2\pi T D_x$ for $D$ meson (left plot) and $\bar B$ meson (right plot) (taken from Ref.~\cite{Tolos:2013kva,Torres-Rincon:2014ffa}). For recent updates in the high-temperature phase for $D$ meson, see Refs.~\cite{Ozvenchuk:2014rpa,Berrehrah:2014tva}. }
\label{fig:diffusion}
\end{center}
\end{figure}

One interesting observable is the behaviour of the spatial diffusion coefficient $D_x$ that appears in Fick's diffusion law in  medium, given in terms of the scalar $F(p)$ and $\Gamma(p)$ coefficients \cite{Tolos:2013kva,Torres-Rincon:2014ffa}. In Fig.~\ref{fig:diffusion}  we show  $2\pi T D_x$  for $D$ and $\bar B$ mesons following isentropic trajectories ($s/n_B$=ct) from RHIC to FAIR energies. For the $D$ mesons,  we observe that the dependence of the $2\pi T D_x$ on the entropy per baryon is similar in the hadronic  and quark phase (below and above the transition temperature of $T_c \sim 150 {\rm MeV}$, respectively). The possible matching between curves in both phases for a
given $s/n_B$  seems to indicate the possible existence of a minimum in the $2\pi T D_x$ at the phase transition \cite{Ozvenchuk:2014rpa,Berrehrah:2014tva}.

\section*{Acknowledgments}
C.~H.-D. thanks the support of the JAE-CSIC Program. L.T. acknowledges support from the Ram\'on y
Cajal Research Programme from Ministerio de Ciencia e Innovaci\'on and from
FP7-PEOPLE-2011-CIG under Contract No. PCIG09-GA-2011-291679.
This research was supported by  Spanish Ministerio de Econom\'\i a y Competitividad and
European FEDER funds  under contracts FPA2010-16963, FIS2011-28853-C02-02, FPA2013-43425-P,
FIS2014-59386-P,  FIS2014-51948-C2-1-P and 
FIS2014-57026-REDT, and Junta de Andaluc{\'\i}a grant FQM-225.

\end{document}